\documentclass[a4paper,fleqn]{cas-sc}

\usepackage[numbers]{natbib}
\usepackage{svg}
\usepackage{xurl}
\usepackage{hyperref}
\usepackage{cleveref}
\usepackage{siunitx}
\usepackage{float}
\usepackage{comment}
\usepackage{enumitem}
\usepackage{booktabs}

\DeclareSIUnit{\rpm}{\revolution\per\minute}

\crefname{figure}{Figure}{Figures}       % singular, plural
\Crefname{figure}{Figure}{Figures}       % am Satzanfang
\crefname{section}{Section}{Sections}
\Crefname{section}{Section}{Sections}

\def\tsc#1{\csdef{#1}{\textsc{\lowercase{#1}}\xspace}}
\tsc{WGM}
\tsc{QE}
\begin{document}
\let\WriteBookmarks\relax
\providecommand{\path}[1]{#1}

\shorttitle{}    

\shortauthors{Hemmerich et al.}  

\title [mode = title]{Unified Embodiment Description for functional evaluation of used components in circular manufacturing systems}  

\author[1]{Jonas Hemmerich}
\credit{Conceptualization, Methodology, Investigation, Formal analysis, Visualization, Writing -- original draft, Writing -- review \& editing}
\cormark[1]

\author[2]{Dominik Koch} %% Author name
\credit{Investigation, Formal analysis, Visualization, Writing -- original draft}
\author[1]{Victor Mas} %% Author name
\credit{Conceptualization, Writing -- original draft}
\author[1]{Nehal Afifi} %% Author name
\credit{Conceptualization, Writing -- original draft}
\author[2]{Edwin Blum} %% Author name
\credit{Investigation, Formal analysis}

\author[2]{Gisela Lanza} %% Author name
\credit{Supervision, Writing -- review \& editing}
\author[1]{Sven Matthiesen} %% Author name    
\credit{Supervision, Writing -- review \& editing}
\author[1]{Patric Grauberger} %% Author name
\credit{Funding acquisition, Supervision, Project administration, Conceptualization, Writing -- original draft}

\affiliation[1]{organization={IPEK - Institute of Product Engineering, Karlsruhe Institute of Technology (KIT)},%Department and Organization
            addressline={Kaiserstr. 10}, 
            city={Karlsruhe},
            postcode={76131}, 
            state={Baden Wuerttemberg},
            country={Germany}}

\affiliation[2]{organization={wbk - Institute of Production Science, Karlsruhe Institute of Technology (KIT)},%Department and Organization
            addressline={Kaiserstr. 12}, 
            city={Karlsruhe},
            postcode={76131}, 
            state={Baden Wuerttemberg},
            country={Germany}}
            
\cortext[cor1]{Corresponding author.\newline\mbox{\hspace*{2.35em}}E-mail address: \href{mailto:jonas.hemmerich@kit.edu}{jonas.hemmerich@kit.edu} (J. Hemmerich)}


\begin{abstract}
Circular manufacturing systems require functional evaluation of used components based on their physical state.
Existing approaches describe this state from separate perspectives, such as design, manufacturing, and degradation, resulting in fragmented and incompatible representations.
As a consequence, the physical state cannot be reliably linked to the functional behavior of the corresponding subsystem, which is a prerequisite for informed R-strategy decisions.
This paper introduces the Unified Embodiment Description (UED), a state-dependent representation of mechanical components structured into two coupled layers.
The first layer is a unified characteristic space, which adapts and extends as new lifecycle effects emerge.
The second layer consists of functionally derived tolerance regions that link these embodiment characteristics to the functional behavior of the surrounding subsystem.
A supporting UED method guides the model-building process of both layers.
The UED is demonstrated in a case study on the spindle shaft of an angle grinder, in which manufacturing variations and degradation patterns such as polishing wear and scratches are quantified.
These embodiment changes are embedded into the unified characteristic space and translated into functionally derived tolerance regions through experimental testing of the spindle-bearing subsystem.
The results show that embodiment changes induced over the lifecycle can be consistently integrated within the unified characteristic space and that the relations between embodiment and functional behavior can be quantified to support end of life decisions. Overall, the UED provides a foundation for embodiment modeling that adapts to component state and enables decision-making based on functional evaluation for used components in circular manufacturing systems.
\end{abstract}

\begin{graphicalabstract}
\includegraphics[width=\textwidth]{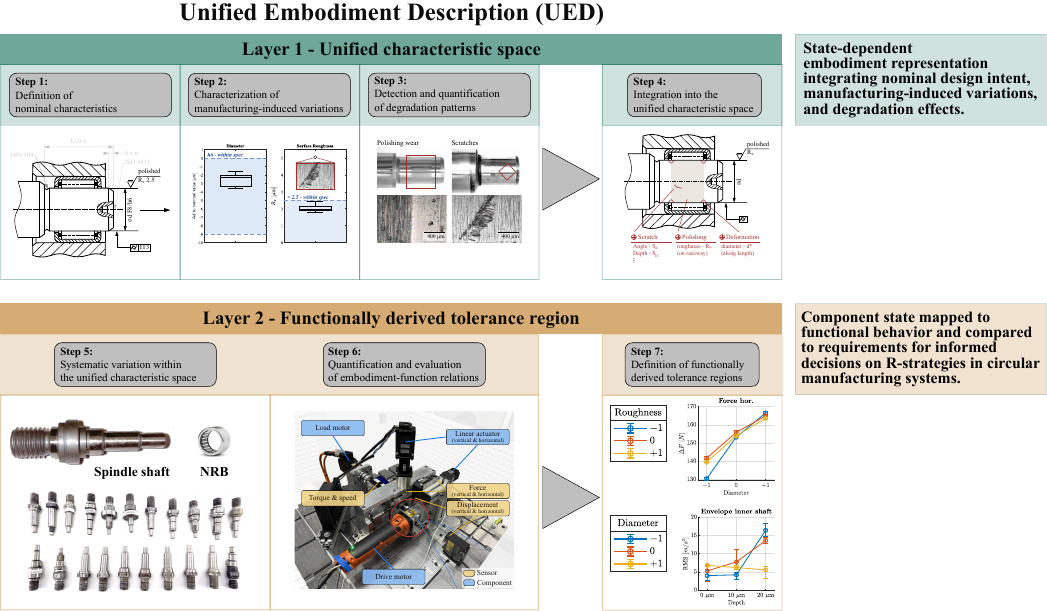}
\end{graphicalabstract}

\begin{highlights}
\item A two-layered Unified Embodiment Description models lifecycle-induced state changes.
\item Layer 1 unifies nominal design, manufacturing variation, and degradation effects.
\item Layer 2 derives tolerance regions for decision-making in circular manufacturing.
\item A supporting method identifies and evaluates functionally relevant characteristics.
\item A spindle shaft case study demonstrates the approach under real-world conditions.
\item Design changes, wear, and scratches are linked to subsystem stiffness and vibration.
\end{highlights}

% Keywords
\begin{keywords}
Circular manufacturing systems \sep Remanufacturing \sep Degradation modeling \sep Condition monitoring \sep Functional evaluation
\end{keywords}

\maketitle

% Main text
\section{Introduction}
\label{sec:introduction}
% Circular Economy
Sustainability has become a central objective in production and product development, with the circular economy emerging as a key concept to decouple economic growth from resource consumption.
Within this context, various R-strategies such as repairing, remanufacturing, and recycling enable different pathways for extending product lifecycles or recovering materials.
However, these strategies involve an inherent trade-off between retaining the value embodied in used products, such as repair or remanufacturing, and enabling innovation through future product development ~\citep{parkerRemanufacturingMarketStudy2015, CircularEconomyBusiness2015}.
\\
% Circular Factory
To reconcile this trade-off, the concept of the circular factory has been proposed, aiming to combine value retention with product innovation through the reuse of components and subsystems in new product generations~\citep{lanzaVisionCircularFactory2024}.
In the circular factory, reused products are no longer evaluated solely against their original specifications but must satisfy evolving functional requirements.
This shifts the perspective on the trade-off between value retention and innovation toward the question of whether the available component remains capable of fulfilling the required functional behavior under changing conditions~\citep{graubergerEnablingVisionPerpetual2024}.
\\
% Decision making
Existing approaches for evaluating R-strategies commonly rely on criteria such as cost, time, resource consumption, or emissions~\citep{theiligLifeCycleAssessment2024,duIntegratedMethodEvaluating2012}. 
While these criteria are essential for evaluating economic and environmental sustainability, the system performance is typically only indirectly captured via indicators such as quality or feasibility~\citep{jiangDevelopmentMulticriteriaDecision2011}. 
However, ensuring that reused components and subsystems fulfill functional requirements necessitates an explicit evaluation of functional behavior. 
Functional behavior, in turn, is not directly specified but emerges from the embodiment of a technical system, i.e., geometry and material, and its interaction with the environment, such as contact conditions or applied loads~\citep{geroSituatedFunctionBehaviour2004,weberModellingProductsProduct2014,graubergerContactChannelApproach2020}.
Consequently, decision-making in the circular factory relies on the evaluation of embodiment and its relation to the resulting functional behavior.
\\
% Embodiment description
In contrast to conventional design scenarios, the embodiment of reused components does not correspond to a well-defined state. 
Instead, it results from the superposition of multiple influences originating across the product lifecycle, which converge within the circular factory:
(1) Degradation mechanisms accumulated during prior use alter the embodiment of incoming components in an instance-specific way.
(2) Intentional design changes in product development modify the nominal embodiment to adapt used components to new requirements;
(3) Manufacturing processes in both linear production and reprocessing cause deviations from the intended embodiment.
These influences interact in complex and potentially conflicting ways.
As a result, evaluating functional behavior becomes more challenging, as it requires a consistent description of heterogeneous and evolving embodiment states.
\\
% Current approaches
Current approaches in engineering design, manufacturing, and degradation analysis address this challenge only in a fragmented manner, as they describe embodiment states from domain-specific perspectives.
Design methods primarily focus on the initial nominal state of a product, providing models that link embodiment to expected functional behavior~\citep{pahlEngineeringDesign2007}.
In contrast, quality management mainly captures manufacturing-induced deviations from nominal specifications~\citep{soderbergDigitalTwinRealtime2017}. 
Their evaluation is typically centered on process capability or dimensional conformity rather than on resulting functional behavior.
An integration of design and manufacturing perspectives can be found in tolerance analysis and uncertainty-based design~\citep{morseTolerancingManagingUncertainty2018}. 
However, although changes in embodiment are recognized as a source of uncertainty throughout the product lifecycle, its functional implications are commonly assessed only relative to the initial nominal design state.
Degradation is specifically addressed in condition monitoring and predictive maintenance, where emphasis is placed on the temporal evolution of system states and the detection of failure.
These approaches predominantly infer degradation through changes in functional behavior signals~\citep{dadfarniaComprehensiveEvaluationsCondition2025}, while the embodiment of the degraded state receives comparatively little attention.
\\
% Gap and Research Question
%Existing work addresses embodiment, degradation, and manufacturing deviations separately, but lacks a unified representation that integrates these aspects for lifecycle-consistent decision support.
While these approaches provide valuable insights, their domain-specific focus prevents a coherent description of embodiment that combines nominal design intent, manufacturing-induced variations, and degradation effects. 
Consequently, the central problem is that no systematic basis exists for decision-making, as embodiment states cannot be consistently compared with respect to their functional behavior.
Key decisions in the circular factory remain unresolved, such as whether a degraded component can still be reused, to what extent intended design changes can compensate degradation, or how these factors jointly affect functional behavior.
This leads to the following research question:
\textit{How can the superimposed embodiment of a component be described in a way that enables systematic quantification of its functional behavior for decision-making in circular manufacturing systems?}
\\
% Contribution
To address the research question, this paper introduces the Unified Embodiment Description (UED) as a state-dependent representation of mechanical components based on the superposition of functionally relevant embodiment characteristics. 
Structured into two coupled layers, it combines a unified characteristic space that captures lifecycle-induced embodiment changes with functionally derived tolerance regions that link these characteristics to the functional behavior of the corresponding subsystem. 
\\
The main contributions of this paper are as follows:
\begin{enumerate}
    \item A UED for representing lifecycle-induced state changes of mechanical components and supporting decision-making through comparison of achievable functional behavior with corresponding requirements.
    \item A supporting UED method that guides the model-building process by identifying and evaluating functionally relevant embodiment characteristics arising from design intent, manufacturing-induced deviations, and degradation effects.
    \item An experimental case study on the spindle shaft of an angle grinder demonstrating the construction of the two UED layers under real-world conditions.
\end{enumerate}
% Paper Structure
The remainder of this paper is structured as follows.
\cref{sec:related_work} reviews existing approaches for describing embodiment and its relation to functional behavior.
\cref{sec:proposed_method} presents the conceptual basis of the UED and the procedure of its supporting method.
\cref{sec:overview_casestudy} reports the results from the experimental case study on the spindle shaft.
\cref{sec:discussion} discusses the concept of the UED and its implications for circular manufacturing, while \cref{sec:conclusion} concludes the paper and outlines directions for future research.

\section{Related Work}
\label{sec:related_work}
This section reviews existing approaches for describing embodiment and its relation to functional behavior.
Embodiment is understood as the physical realization of a technical system, encompassing both its geometry and material attributes~\citep{pahlEngineeringDesign2007}. 
The concept of behavior is formalized in the Function–Behavior–Structure (FBS) framework~\citep{geroSituatedFunctionBehaviour2004}, while the Characteristics Property Modeling (CPM) framework~\citep{weberModellingProductsProduct2014} further specifies functionally relevant aspects of product behavior.
Based on these conceptual foundations, the related work is structured along three perspectives: design-oriented representations of nominal embodiment, manufacturing-related approaches addressing stochastic variations, and degradation-focused methods capturing time-dependent changes in component condition. 
\subsection{Description of embodiment}
In \textbf{engineering design} embodiment constitutes the tangible realization of design decisions and provides the basis from which a component's functional behavior emerges. 
A widely used approach to describe the nominal embodiment is through characteristics, which capture geometric and material properties in a structured manner~\citep{weberModellingProductsProduct2014}. 
The description using such characteristics is guided by established standards and norms, such as the Geometrical Product Specifications (GPS) \citep{DINISO1101}. 
These specifications define nominal values and tolerance ranges for dimensions, form, and positional relationships in technical drawings. 
Furthermore, the specifications allow the nominal embodiment to be fully represented in digital computer-aided design environments, enabling precise documentation and evaluation. 
While this approach provides a clear specification of the intended design, it is limited to the ideal component state including nominal geometry and tolerance definitions. 
Consequently, classical descriptions are insufficient to evaluate embodiment in scenarios involving deviations caused by manufacturing or degradation, as encountered in circular manufacturing contexts.
\\
In industrial practice, \textbf{manufacturing}-induced variations are addressed through quality management and tolerance analysis. 
For linear manufacturing, product quality is typically ensured through structured quality assurance processes, in which components are verified against defined specifications either by sampling or 100\% inspection~\citep{juran2016}. 
While these methods are effective for detecting non-conformities and ensuring production quality, they primarily operate as verification procedures against nominal specifications and do not provide a detailed representation of the actual as-manufactured geometry.
To address these limitations, tolerancing research provides approaches for representing and analyzing geometric variations beyond nominal specifications. 
Within tolerancing, the GeoSpelling framework is of particular interest, as it builds directly on the classical GPS concepts by incorporating so called skin models~\citep{dantanGeometricalProductSpecifications2008}. 
Skin models represent the real physical surface of a component, including both ideal and non-ideal geometric features resulting from manufacturing processes. 
Based on the skin models, deviations of geometric embodiment characteristics can be derived and used for tolerance analysis, virtual assembly, and evaluation of geometric variability~\citep{schleichSkinModelShapes2014}. 
GeoSpelling has also been applied to characterize real components, including cases where degradation effects are already present~\citep{zhangGeometricProductSpecification2015}, which highlights its relevance for circular manufacturing. 
Overall, this approach enables a more realistic representation of geometric variation beyond nominal specifications, with direct implications for characteristic tolerancing.
Despite improving the representation of deviations, tolerancing approaches such as GeoSpelling do not explicitly account for the dynamic evolution of embodiment over the product lifecycle.
As a result, geometric variation is primarily evaluated with respect to nominal design states, while its implications for functional behavior in reprocessing scenarios remain only partially considered.
\\
In circular manufacturing, \textbf{Degradation} represents a key source of embodiment variability that cannot be captured by fixed inspection assumptions, requiring flexible and adaptive evaluation strategies~\citep{kochEnhancingVisualInspection2025}. 
To capture and evaluate the condition of returned components, vision-based sensor systems combined with image processing techniques have been widely adopted~\citep{yeReviewMachineVisionBased2016}.
These approaches enable detection and classification of surface defects such as fatigue, wear, and cracks~\citep{tabernikSegmentationbasedDeeplearningApproach2020,liuBevelGearQuality2016}, as well as quantitative characterization using 2D and 3D measurement data providing the basis for data-driven reprocessing decisions~\citep{kaiserAdaptiveAcquisitionPlanning2025}. 
\\
Recent advances in machine learning further enable the automated extraction of quantitative defect descriptors directly from image data, reducing reliance on manual evaluation and supporting scalable inspection workflows. 
These approaches operate on localized image information and generate geometric descriptors of affected embodiment regions.
For example, Wu et al.~\cite{wuSCESSurfaceCondition2026} demonstrate the automated measurement of scratch length using segmentation-based methods, achieving measurement accuracies of over $85\si{\percent}$.
Zou et al.~\cite{zouExperimentalNumericalStudy2023} investigate fretting wear-fatigue in press-fitted axles using microscopy images, quantifying degradation through characteristics such as wear depth, scratch length, and orientation. 
Similarly, Van Maele et al.~\cite{vanmaeleVisionassistedConditionMonitoring2025} describe local pitting damage on helical gears using geometric features such as position, size, and shape. 
\\
Still, current degradation approaches remain limited in their ability to provide a unified representation of degradation effects. 
Most methods focus on defect detection or classification rather than a holistic geometric description of the affected component. 
Even when quantitative descriptors are available, they are typically case-specific and treated as a set of local discrete observations rather than a continuous modification of the global embodiment. 
\\
Across these three perspectives, embodiment is represented either as an idealized design state, as a deviation from nominal specifications, or as a time-dependent condition derived from observed degradation.
The same holds when considering functional behavior as a reference for evaluating these embodiment representations.

\subsection{Relation between embodiment and functional behavior}
The relation between embodiment and functional behavior has been addressed at different levels of abstraction in \textbf{engineering design}.
Conceptual frameworks such as the FBS framework \cite{geroSituatedFunctionBehaviour2004} provide a theoretical foundation by distinguishing function, behavior, and structure as core elements of product design. 
Extensions of this framework incorporate lifecycle-related aspects, such as manufacturability and upgradability \citep{sandersonFunctionBehaviourStructureDesignMethodology2019,umedaDevelopmentDesignMethodology2005} or Design for Upgrade and Remanufacturing guidelines \citep{wuCustomizedDesignMethod2023}. 
While these models extend the scope toward lifecycle considerations, they remain primarily conceptual and do not provide a detailed representation of geometric embodiment or functional behavior.
\\
More detailed approaches are found in engineering design practice, where functional behavior is analyzed through parameter-based models. 
Within the embodiment–function relation, \textbf{manufacturing}-induced variations are generally not explicitly represented as a distinct class of embodiment characteristics affecting functional behavior, but are instead accounted for implicitly during the design phase through tolerancing and design-for-manufacturing strategies.
In this context, sensitivity analysis in conceptual design and optimization techniques in detailed design are used to investigate how intended variations in embodiment characteristics influence functional behavior.
These methods are commonly implemented using Design of Experiments (DoE), which enables a systematic variation of parameters and supports the identification of significant factors \citep{montgomeryDesignAnalysisExperiments2019}. 
The resulting functional behavior is typically evaluated by testing \citep{taheraTestingIncrementalDesign2019}, for example in simulation studies \citep{jingliuSimulationAnalysisBall2023} or in experimental settings \citep{krupkaEffectRealLongitudinal2010}. 
Simulations include physics-based models such as finite element analysis and multibody dynamics, but rely on idealized embodiment representations.
To allow a more detailed representation of embodiment variations across scales, co-simulation approaches and multi-scale modeling techniques enable the coupling of different physical domains and abstraction levels. 
Experimental setups operate on physical components and therefore inherently include actual embodiment states, including manufacturing-induced variations.
Building on nominal descriptions, robust design approaches explicitly account for the effect of embodiment variation on functional behavior by aiming to reduce sensitivity regarding manufacturing processes and operational uncertainties \citep{parkRobustDesignOverview2006}. 
While both approaches enable a quantitative mapping between embodiment variations and functional performance, they are generally restricted to well-defined nominal geometries or controlled manufacturing deviations.
\\
In contrast, \textbf{degradation} effects are primarily addressed in condition monitoring and predictive maintenance approaches, which focus on monitoring the functional behavior of components during operation \citep{fengReviewVibrationbasedGear2023}. 
Sensor-based methods analyze signals such as vibrations or displacements to detect damage and estimate remaining useful life \citep{liIntelligentFaultIdentification2021}. 
While these approaches capture the temporal evolution of system behavior, they provide only an indirect representation of the underlying embodiment. 
Degradation is typically inferred from changes in measured signals, whereas the corresponding embodiment changes are not explicitly modeled or quantified. 
Even when inspection data are used to relate embodiment measurements to physical degradation, this mapping is often non-systematic and strongly dependent on operating conditions and observed damage patterns \citep{zouExperimentalNumericalStudy2023}. 
To enable controlled analysis of degradation effects, some studies introduce artificial damage to investigate its influence on functional behavior under defined conditions \citep{haderleVibrationAnalysisEarly2024}. 
Artificial damage allows for isolating specific degradation mechanisms and studying their effect on system behavior in a controlled experimental setting.
\\
Overall, existing approaches establish conceptual relations between embodiment and functional behavior, quantify the influence of nominal geometric parameters, and observe functional behavior under operational conditions. 
These approaches can, in principle, account for variations arising from design, manufacturing, and degradation perspectives. 
However, they rely on different and often incompatible representations of embodiment states, which limits their ability to serve as a consistent input for systematic analysis of functional behavior across lifecycle stages.

\section{Proposed approach}
\label{sec:proposed_method}
The fragmented representation of embodiment states and its relation to functional behavior motivates a structured methodological approach for consistent decision-making in circular manufacturing contexts.
This section introduces the UED as a state-dependent representation that integrates design, manufacturing, and degradation perspectives into a single characteristic-based model.
A supporting UED method is then presented to guide the model-building process for a given component.
The UED is described according to established guidelines for structured engineering descriptions using the four elements of intended use, core, representation, and procedure \citep{gerickeWhatWeNeed2017}.

\subsection{Unified Embodiment Description}
\begin{figure}[pos=t]
    \centering
    \includegraphics[width=15.4cm]{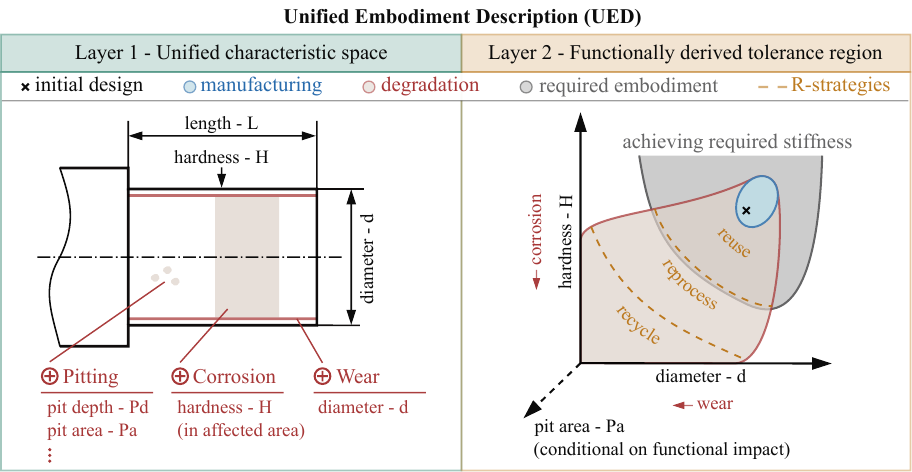}
    \caption{ Layer 1 shows nominal design characteristics (black) and degradation-induced descriptors (red). Layer 2 adds manufacturing variations (blue), required embodiment space (grey), and R-strategy boundaries (orange), arranged by increasing intervention depth.}
    \label{fig:UED_model}
\end{figure}

The \textbf{intended use} of the UED is to support decision-making for returned mechanical components in circular manufacturing systems, where the embodiment state deviates from nominal design.
It enables the assessment of whether a component can be reused, requires reprocessing, or should be recycled by linking observed embodiment states to functional requirements under evolving characteristic conditions.
The UED is applied during the development phase of the next product generation, prior to factory operation, by engineers with existing design knowledge of the technical system.
\\
The \textbf{core} of the UED is based on the principle that all lifecycle-induced changes in mechanical components can be consistently captured through the definition of embodiment characteristics.
This enables a coherent and comparable description of component states across lifecycle stages and supports the systematic analysis of their impact on functional behavior.
\\
On a structural level, the \textbf{representation} of the UED consists of two coupled layers, a unified characteristic space and functionally derived tolerance regions defined within this space.
The two layers represent a decoupling between the embodiment characteristics themselves and their corresponding tolerances. 
This separation is necessary because functional requirements may vary across product generations and usage conditions, while degradation can alter the relevance and interaction of embodiment characteristics over time, making fixed characteristic-specific tolerance definitions insufficient to capture evolving functional constraints.
\Cref{fig:UED_model} illustrates the layer structure for a simplified shaft, which serves as an example throughout this section.
\\
The first layer is the unified characteristic space, which represents the embodiment state of a component. 
Nominal design definitions provide the initial reference state and degradation effects introduce additional or modified characteristics reflecting usage-dependent changes. 
As a result, the unified characteristic space is not static but adapts and extends depending on the observed embodiment state.
Existing embodiment characteristics may remain valid, require adaptation, or be complemented by additional descriptors when new degradation patterns occur.
Manufacturing-induced variation acts on existing nominal characteristics rather than introducing new ones. 
It is therefore not represented in the unified characteristic space but in the baseline distribution against which degradation-induced deviations are interpreted.
In the illustrated example, the nominal design defines the initial characteristics of the shaft, including diameter, length, and surface hardness. 
These characteristics are subject to degradation patterns observed during use. 
Abrasive wear reduces the diameter, corrosion decreases the surface hardness at affected areas, and pitting introduces additional descriptors such as pit depth and pit area that extend the characteristic space.
\\
The second layer of the UED defines functionally derived tolerances within this characteristic space. 
Tolerance limits are not defined as independent bounds of individual characteristics, but as regions in the unified characteristic space that satisfy functional requirements under multivariate interaction. 
These regions are obtained by evaluating combinations of characteristics with respect to functional behavior, resulting in system-level constraints rather than isolated characteristic limits.
In the shown example, the functional tolerance region is derived from a required stiffness and spans the two nominal characteristics of diameter and surface hardness.
The resulting region is not axis-parallel but reflects the multivariate interaction of the characteristics. 
Degradation caused by wear and corrosion shift the embodiment state from the manufacturing-induced baseline toward the functional boundary, where the position within or outside the region supports the assessment of suitable R-strategies such as reuse, reprocessing, or recycling of a used component.
If further investigation of the pit area yields an impact on the shaft stiffness, additional dimensions need to be considered in the characteristic space.
\\
The assignment of R-strategies follows from the position of the current embodiment state relative to the functional tolerance region. 
The distance to the manufacturing-induced variation quantifies the degradation accumulated through use, while the position relative to the tolerance region determines whether the required functional behavior is still satisfied. 
Combined with the available reprocessing operations and their achievable shifts in the characteristic space, these relations define the admissible strategy. 
Embodiment states within the tolerance region permits direct reuse of the components. 
A state at the boundary or outside this region but returnable to a functionally required position through operations such as grinding or re-hardening supports reprocessing. 
States beyond the reach of any reprocessing operation are assigned to recycling, preserving only the material value.
\\
The \textbf{procedure} for constructing the two layers is based on the supporting UED method.

\subsection{UED method}
\begin{figure}[pos=t]
    \centering
    \includegraphics[width=14.5cm]{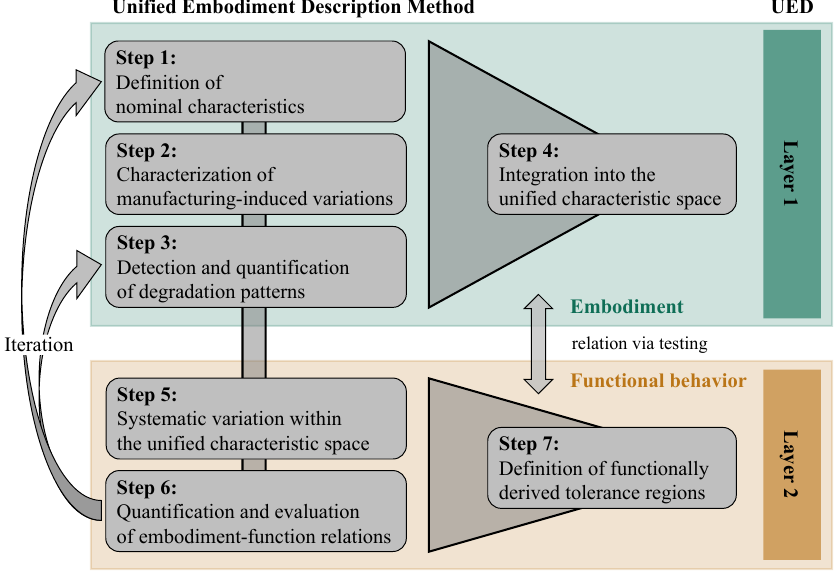}
    \caption{Seven-step methodology for constructing the UED. Steps 1–4 establish the unified characteristic space (layer 1) by integrating design and degradation information. Steps 5–7 define the functionally derived tolerance region (layer 2) by linking embodiment characteristics to functional behavior, and can be revisited iteratively as new insights are gained.}
    \label{fig:proposed_method}
\end{figure}

The UED method is illustrated in \cref{fig:proposed_method} and follows seven steps, of which the first four establish the unified characteristic space (layer 1) and the last three define the functionally derived tolerance regions (layer 2). 
First, nominal embodiment characteristics are identified based on functionally relevant GPS from the design definition.
Second, manufacturing-induced variations are characterized to establish a reference distribution of as-manufactured components. 
Third, degradation effects are detected and quantified using appropriate descriptors.
Fourth, these embodiment descriptors are integrated into the unified characteristic space that consistently represents the individual component state. 
Fifth, systematic variation of embodiment characteristics is performed to explore their influence on functional behavior. 
Sixth, the impact on functional behavior is quantified and evaluated based on testing. 
Seventh, functionally derived tolerance regions are established based on functional requirements.
The method can be revisited iteratively as the evaluation gains new insights about functional relevance causing nominal characteristics or degradation descriptors to be excluded, adapted, or added.
 
\subsubsection{Step 1: Definition of nominal characteristics}

%Nominal design values and associated drawing tolerances are extracted as input information from technical drawings. While nominal values define the initial embodiment characteristics, tolerances are not directly used within the unified characteristic space but serve as reference bounds for subsequent variation in Step 5.

The first step identifies functionally relevant nominal characteristics based on classical GPS. 
It provides a reduced and structured representation of the component geometry in its ideal design state, serving as the initial reference for all subsequent analyses.
In circular manufacturing contexts, these nominal characteristics additionally serve as a cross-manufacturer and cross-generational reference for comparing geometrically similar components from different product instances.
The step begins by identifying embodiment characteristics that describe the geometry in a measurable and unambiguous way, such as dimensions, form deviations, or surface-related properties. 
Extraction of the characteristics heavily relies on the technical documentation from the previous design phase of the returned product, including technical drawings and associated specification data.
In addition, tolerance specifications and constraints associated with each selected characteristic are incorporated to define the allowable variation in linear production.
Since each characteristic introduces an additional dimension, a subsequent reduction is performed to retain only those characteristics that are assumed to be relevant for functional behavior within the respective system context. 
This reduction is guided by defining relevant aspects of the functional behavior that are influenced by the component (e.g., vibration, noise, heat generation, sealing, efficiency).
\\
The result of the first step is a set of nominal characteristic that represents the idealized embodiment of the component, including its functional characteristics and associated tolerance definitions. 
This set serves as the reference domain against which real component states in circular manufacturing contexts are interpreted.
While the characteristics populate the unified characteristic space (layer 1), the associated nominal tolerances are not directly applied as constraints. 
Instead, they serve as a baseline for interpreting manufacturing-induced variation in step 2 and inform the derivation of functionally derived tolerance regions in step 7.

\subsubsection{Step 2: Characterization of manufacturing-induced variations}
For returned components, manufacturing-induced variations and degradation-induced changes cannot be strictly separated in the observed geometry, as both contribute to the same measured embodiment state. 
Consequently, degradation is treated as the primary source of state variation in circular manufacturing contexts, while manufacturing-induced deviations define the reference framework for interpreting geometric measurements.
In addition, geometric specifications such as tolerances for surface- or form-related characteristics are often asymmetric or only partially constraining. 
As a result, measured values cannot be directly interpreted in isolation, as their position within the intended design space is not uniquely defined. 
A reference-based model is therefore required to enable consistent interpretation of geometric states.
\\
Starting from the nominal characteristic space defined in the first step, real manufacturing data based on process monitoring or quality control are used to populate the characteristic space with observed variations.
For each characteristic, its statistical distribution is analyzed with respect to bias within the tolerance range and overall scatter. 
As a results systematic tendencies, such as shifts toward tolerance limits, as well as stochastic process-related variation are revealed.
This statistically grounded baseline distribution supports the interpretation of returned components such as presence and severity of degradation patterns.
\\
The outcome of this step is a manufacturing-based reference model of as-manufactured components that extends the nominal characteristic space with empirical information on manufacturing capability.

\subsubsection{Step 3: Detection and quantification of degradation patterns}
The third step detects and quantifies occurring degradation patterns in used components by first establishing a structured knowledge base.
As this step relies on empirical measurements, it requires used components exhibiting representative degradation.
Similar to condition monitoring, the analysis is driven by individual patterns, while being consistently expressed through the use of embodiment characteristics.
\\
%\textbf{Definition of the product-specific degradation knowledge base}\\
The definition of the degradation knowledge base is guided by standards for machine elements, domain expertise, and empirical observations of returned components, which provide information on typical probability and severity.
Based on the application context and operating conditions of the component, relevant degradation patterns are selected and structured. 
For each pattern, the corresponding embodiment manifestation is defined in terms of observable geometric effects. 
From this, measurable features and characteristics for detection and quantitative descriptors are derived.
Degradation patterns are distinguished into continuous effects, which affect the overall embodiment (e.g., abrasive wear), and discrete effects, which are localized in space (e.g., pitting). 
The resulting knowledge base links degradation patterns, their geometric manifestations, and associated features in a structured form.
\\
%\textbf{Detection and quantification of degradation patterns}\\
Based on the knowledge base, degradation patterns are detected and quantified using appropriate measurement strategies. 
The selection of a suitable detection strategy depends on the degradation manifestation, functional impact, and accessibility of the component and may include geometry-based (e.g., optical, tactile) or behavior-based (e.g., vibration, friction) approaches.
Automated detection approaches are particularly relevant for discrete degradation patterns requiring spatial localization and for components with limited accessibility.
\\
Quantification is performed using the characteristics defined in the knowledge base. 
Depending on the required descriptor type and measurement resolution, different measurement strategies and sensing systems may be required.
The resulting measurements are evaluated against the manufacturing-based reference model established in step 2 to assess whether observed deviations exceed expected manufacturing variability and can be attributed to degradation-induced deviations.
\\
The outcome of this step is a structured description of degradation patterns expressed in terms of geometric characteristics and corresponding measurement strategies. 

\subsubsection{Step 4: Integration into the unified characteristic space}
The fourth step integrates the outputs of steps one to three into a consistent and extendable embodiment description of the individual returned component. 
The nominal characteristic space from step 1 provides the structural reference, the manufacturing model from step 2 supports interpretation of degradation, and the quantified degradation patterns from step 3 capture state-dependent changes.
Depending on the detected degradation patterns, the nominal state in the unified characteristic space is selectively adapted or extended. 
In this sense, the integration follows a qualitative activation logic, where additional descriptors are introduced only when the corresponding degradation patterns are present.
The introduction of characteristics comprises two activities.
%\textbf{Initial comparison of characteristics}\\
First, the descriptors of each detected degradation pattern are compared with the nominal characteristic space. 
Depending on their relation to the existing characteristics, three cases are distinguished:
\begin{itemize}
    \item Case 1: Existing characteristics are sufficient.\\
    The degradation pattern can be expressed using an existing characteristic without modification \\
    (e.g., change in diameter due to wear).
    \item Case 2: Existing characteristics require adaptation.\\
    The characteristic remains relevant, but its description must be extended or reformulated \\
    (e.g., localized evaluation of hardness due to corrosion).
    \item Case 3: New characteristics are required.\\
    The degradation pattern is not represented in the nominal space and additional characteristics must be introduced 
    (e.g., additional descriptors for pitting).
\end{itemize}
Cases 1 and 2 mainly occur for continuous degradation effects, whereas localized discrete damage often requires additional characteristics according to case 3.
\\
%\textbf{Subsequent identification of interactions}\\
After adaptation of the characteristic space, interactions between characteristics are considered in the second step. 
Newly introduced degradation descriptors may not only have direct effects, but may also modify existing characteristics.
For example, a scratch may be represented by its depth as a new localized characteristic, while simultaneously reducing local diameter and increasing roughness. 
Such dependencies are documented and, where necessary, incorporated into subsequent measurement and evaluation procedures.
\\
The outcome of this step is a structured embodiment description that combines global characteristics with spatially referenced degradation descriptors. While global characteristics can be expressed as scalar values, the spatial localization of discrete degradation patterns benefits from a 3D representation of the component, in which position, orientation, and extent of localized features are unambiguously assigned. This combined representation captures the state-dependent embodiment of real components and provides the basis for the subsequent analysis of functional behavior.

\subsubsection{Step 5: Systematic variation within the unified characteristic space}
After establishing the qualitative unified embodiment description in step 4, its implications for functional behavior must be evaluated, as the relevance of individual characteristics has not yet been quantified. 
As functional behavior emerges from the interaction of components within a technical system, embodiment descriptions alone are not sufficient for decision-making in circular manufacturing contexts. 
\\
The fifth step therefore prepares the systematic evaluation by deriving a structured study design within the unified characteristic space. 
Instead of exhaustively varying all unified embodiment characteristics in a full-factorial manner, the approach follows a hypothesis-driven strategy in which knowledge of the preceding analysis is used to anticipate expected relations, likely interactions, and relevant variation ranges.
Characteristic variation follows two complementary perspectives: controllable variation through design changes or reprocessing measures, and inherent variation caused by degradation effects. 
Representative factor levels are selected to cover observed component states and to enable systematic exploration of the adjacent parameter space.
The systematic investigation of degradation-related embodiment changes requires the introduction of artificial damage to realize the physical embodiment state of specimens for experimental testing.
By expressing degradation-induced and nominal characteristic changes within the same unified description, both domains can be jointly investigated. 
This enables the systematic analysis of main effects, interactions, and potential compensation mechanisms.
For evaluation in the subsequent step, a reference state is defined representing a functionally compliant baseline (e.g., a new or nominally conforming component). 
Measured responses of varied embodiment states are compared against this reference state to determine how embodiment changes influence functional behavior.
\\
The outcome of this step is a study design for evaluating the functional relevance of embodiment characteristics within the unified embodiment description.

\subsubsection{Step 6: Quantification and evaluation of embodiment-function relations}
The sixth step generates and evaluates measurable data describing the functional behavior based on the study design of step 5.
An appropriate testing environment is established, including experimental setups and/or simulation models capable of reproducing the defined characteristic variations while reliably capturing the metrics of the functional behavior.
As this step comes at considerable effort, the definition of functional metrics and establishment of testing environment can often build on already present validation criteria, testing procedures, and performance targets, highlighting the knowledge gained from the original product development process. 
Subsequent data analysis identifies relevant dependencies, sensitivities, and potential interactions between characteristics. 
These relations are summarized in a suitable form, such as empirical models, response surfaces, threshold values, or ranked influence factors, to support interpretation and later application.
\\
The outcome of this step is a reliable dataset linking embodiment variations to quantified functional responses. 
This provides an empirical basis for understanding how changes in embodiment affect system behavior without requiring a fully predictive physical model.

\subsubsection{Step 7: Definition of functionally derived tolerance regions}
The seventh step derives practically applicable characteristic limits and decision rules from the quantified relations between embodiment variation and functional behavior.
For this purpose, the functional requirements of the investigated product variant must be specified. 
Based on these requirements, tolerance regions associated with sufficient functional behavior are identified and linked to the responsible embodiment characteristics. 
Both individual effects and relevant interactions between multiple characteristics are considered.
\\
The resulting tolerance regions are interpreted with regard to suitable actions in circular manufacturing contexts. 
Depending on the observed characteristic state, this may include direct reuse, targeted reprocessing, intensified inspection, or recycling of the component. 
In addition to functional requirements, manufacturing, assembly, and economic constraints should also be considered when defining practically applicable boundaries in accordance to multi-criteria decision models.
\\
The outcome of this step is a set of tolerance regions for relevant embodiment characteristics, complemented by application-oriented decision guidelines. These limits enable the unified embodiment description to support operational decision-making for used components.
Beyond short-term decision support, the identified sensitivities and recurring degradation patterns can also be transferred to future product generations, thereby supporting more robust designs for circular use and closing the loop between embodiment characterization, functional evaluation, and product improvement.

\section{Case study} 
\label{sec:overview_casestudy}
The following case study demonstrates the application of the UED method in a circular manufacturing context for angle grinders, focusing on the reuse and reprocessing of mechanically degraded components. 
A spindle shaft is selected as a representative component to illustrate the method under real world engineering conditions. 
In this case study, the analysis is restricted to geometric embodiment characteristics, while material-related effects are excluded to maintain a focused and measurable application scope.
First, the technical system and case setting are introduced. 
Subsequently, the individual steps of the method are applied and the corresponding results are presented.

\subsection{Case description}
\begin{figure}[pos=t]
    \centering
    \includegraphics[width=14.0cm]{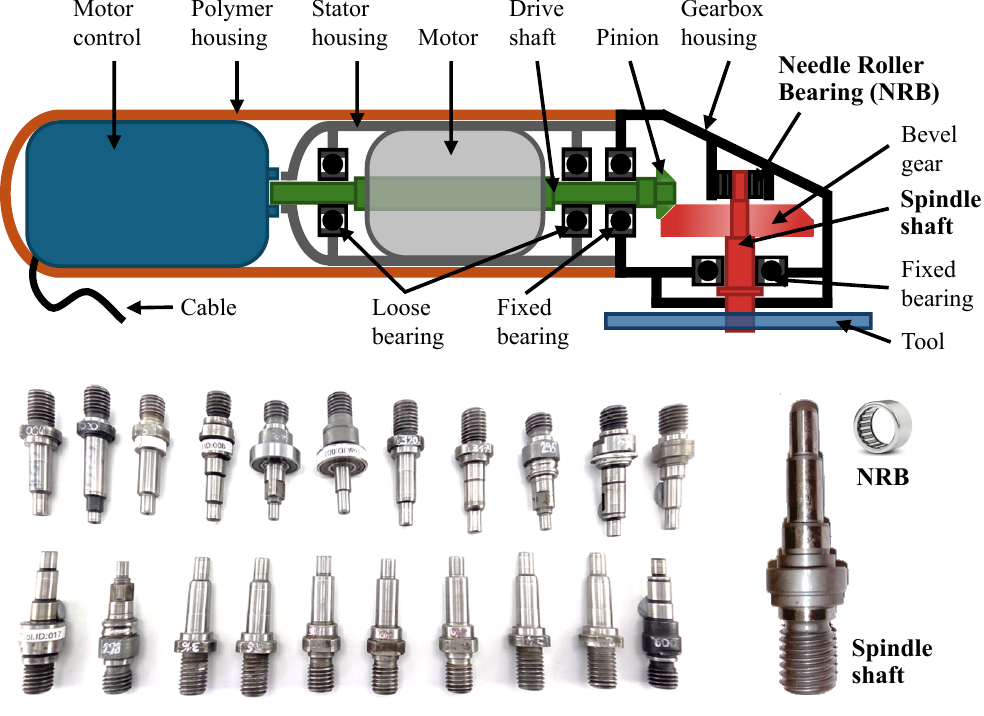}
    \caption{Principal sketch of the \textit{Fein CG15-125BL} angle grinder based on \citep{hemmerichHowDealProducts2025}, showing the product structure with relevant components (top). The bottom view details the spindle shaft, needle roller bearing (NRB), and the full set of analyzed spindle shafts.}
    \label{fig:casestudy_system}
\end{figure}

Angle grinders are portable power tools used for cutting and grinding hard materials such as steel, concrete, or stone. 
\cref{fig:casestudy_system} shows a schematic representation of the technical system and the corresponding component of the case. 
The spindle shaft of the output stage is selected as the investigated component due to its high functional relevance and sensitivity to geometric deviations. 
It transmits the torque of the drivetrain while supporting loads originating from the gear stage and operator-induced forces. 
A needle roller bearing with drawn cup and without inner ring (NRB) supports the spindle shaft in the gearbox housing.
As the needles run directly on the shaft surface the subsystem has a high potential functional sensitivity to embodiment deviations.
\\
From a circular manufacturing perspective, the spindle shaft is particularly relevant because returned angle grinders from different generations and manufacturers often exhibit highly similar subsystem architectures, creating the potential for component reuse beyond the original product instance. 
At the same time, differences in nominal geometry between manufacturers and accumulated degradation from prior use lead to heterogeneous embodiment states. 
This creates a representative circular decision problem, in which the UED is used to determine whether a used spindle shaft can be directly reused, requires reprocessing, or should be recycled, and to what extent nominal design differences interact with degradation effects.
\\
The case study is based on a dataset of 63 spindle shafts, including 57 used and 6 new components originating from 10 different manufacturers.
Results in this paper are restricted to 21 used spindle shafts from five different manufacturers (A-E) shown in \cref{fig:casestudy_system} to account for a sufficient number of samples for valid conclusions. 
For the purpose of cross-manufacturer compatibility in the steps 1-4 and functional evaluation in the steps 5-7, the \textit{Fein CG15-125BL} angle grinder of manufacturer A is used as the reference perspective.
This angle grinder has a nominal diameter of $d=7\,\si{\milli\metre}$ and uses the \textit{SKF HK 0709} as dedicated NRB.
The embodiment is analyzed using standardized metrological procedures for new shafts and adapted measurement strategies for used shafts exhibiting degradation.
To evaluate the influence of embodiment deviations on functional behavior, dedicated test specimens with artificial damage and an experimental test setup are used.

\subsection{Results}
\subsubsection{Step 1: Definition of nominal characteristics}
\label{sec:embodiment_nominalgeometry}
\begin{figure}[pos=t]
    \centering
    \includegraphics[width=16cm]{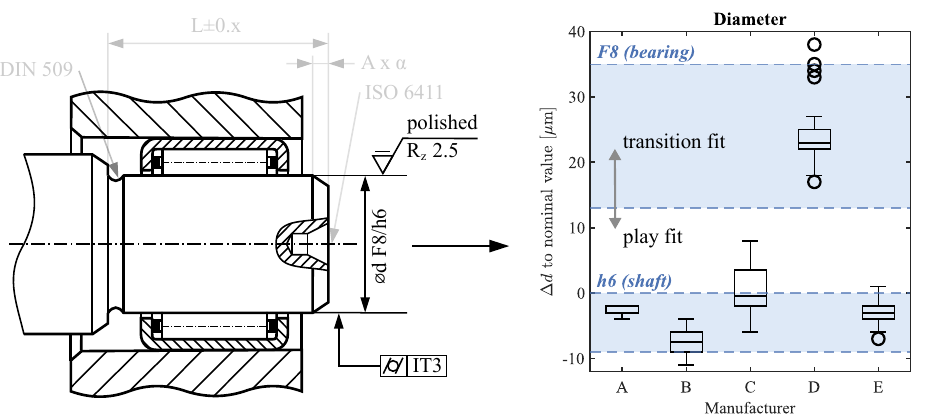}
    %\caption{Technical drawing \citep{norm_roller_iso} including diameter and tolerance zone for different manufacturers}
    \caption{Technical drawing of the spindle-bearing interface (left) and diameter measurements from five manufacturers A-E (right). Based on ISO~3245, relevant embodiment characteristics are extracted, with the measurements highlighting manufacturer-specific fit strategies.}
    \label{fig:embodiment_nominalgeometry}
\end{figure}

The nominal characteristics are derived from GPS defined in the ISO 3245~\citep{norm_roller_iso}.
This norm specifically addresses the design of needle roller bearings with drawn cup and without inner ring.
A corresponding technical drawing of the spindle-bearing subsystem is sketched on the left side of \cref{fig:embodiment_nominalgeometry}. 
Focusing on the contact pair between spindle shaft surface and needles of the NRB, a functional reduction of the characteristic set is performed.
Secondary geometric features in the form of length, chamfer, center hole and relief groove are excluded from the nominal characteristics, as they are not considered dominant for the investigated functional behavior in this case.
The resulting nominal characteristic space consists of the shaft diameter (inside of h6), the shaft cylindricity (inside of T3) and the shaft surface roughness (polished and  below $Rz=2.5\,\si{\micro\metre}$)
In addition to the shaft diameter, the tolerance of the inner diameter of the NRB is defined as F8. 
\\
In the given diameter range the two tolerances relate to the absolute values of $[-9,0]\,\si{\micro\metre}$ (h6) and $[13,35]\,\si{\micro\metre}$ (F8).
The diameter is defined by a two-sided tolerance, while cylindricity and surface roughness are constrained by one-sided limits. 
From a functional perspective, this implies that improved cylindricity and surface roughness have negligible impact on system behavior.
However, for surface roughness reduced friction in the bearing contact may lead to slipping instead of rolling, thereby increase vibration levels.
\\
To investigate cross-manufacturer and cross-generational compatibility, the right side of \cref{fig:embodiment_nominalgeometry} presents the nominal diameter fit strategies applied by the five different manufacturers.
Diameter measurements were performed using a digital outside micrometer (Mitutoyo Digimatic). For each shaft, measurements were taken at two axial positions outside the raceway and two radial orientations offset by 90° per axial position. Each measurement was repeated twice and averaged, resulting in eight individual readings per shaft.
Results show that only manufacturer A dimensioned the shaft according to ISO~3245, while manufacturer B, C and E remain within a play fit regime. 
Manufacturer D far exceeds the nominal h6 tolerance, resulting in a transition fit with respect to the F8 tolerance zone of the NRB.
Also manufacturer D shows significant variation between individual product generations, which suggests that identical components are rarely reused.
This variation in nominal tolerances between manufacturers highlights the importance of evaluating cross-manufacturer compatibility 
It needs to be assessed whether these components still fulfill the functional requirements for the specific product variant of manufacturer A.

\subsubsection{Step 2: Characterization of manufacturing-induced variations}
\label{sec:embodiment_manufactuingdeviations}
\begin{figure}[pos=b]
    \centering
    \includegraphics[width=15.2cm]{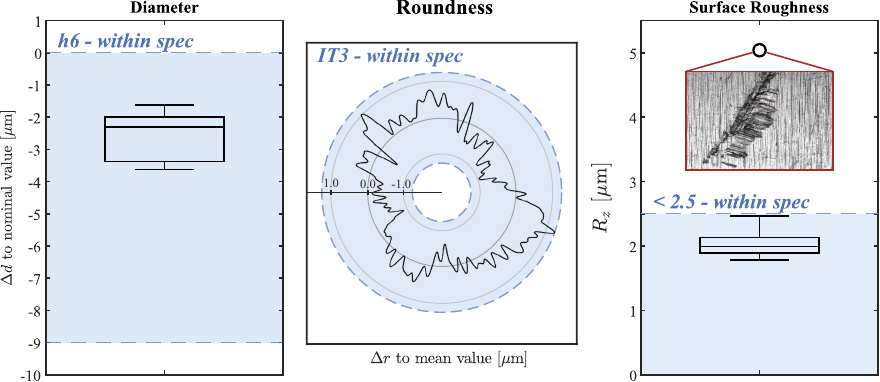}
    \caption{Baseline measurements of diameter (left), roundness (middle), and roughness (right) from manufacturer A. Results shown as box plots against respective specification limits and as a polar profile for roundness. Characteristics are within tolerance for diameter and roundness, with a single outlier in roughness corresponding to localized surface defects.}
    \label{fig:embodiment_manufacturing}
\end{figure}
Based on the nominal characteristic space, manufacturing-induced variations are characterized using measurements from new spindle shafts. 
In this case study, the measurements are carried out manually, due to missing production data.
Six new spindle shafts from manufacturer A are analyzed, as they provide a consistent dataset for the as-manufactured state.
The results of the measurement campaign are shown in \cref{fig:embodiment_manufacturing}.
\\
The measured offsets of the shaft diameter relative to the nominal value are shown on the left side of \cref{fig:embodiment_manufacturing}. 
Diameter measurements were performed using the same digital micrometer and measurement scheme as in Step 1. 
All components lie within the specified h6 tolerance zone. 
A slight systematic offset towards bigger diameters is observed, with a mean offset of $\Delta d = -2.54\,\si{\micro\metre}$ and a maximum offset of $\Delta d = -3.62\,\si{\micro\metre}$. 
The observed standard deviation is $sd = 0.78\,\si{\micro\metre}$.\\
A representative roundness measurement used to assess cylindricity is shown in the center of \cref{fig:embodiment_manufacturing}, obtained with an Alicona \textmu CMM.
For the given nominal diameter, IT3 corresponds to a tolerance of $2.5\,\si{\milli\metre}$. 
This specific measurement and all other measurements remain within this tolerance range, indicating no systematic manufacturing-induced variation in this characteristic.
\\
The right side of \cref{fig:embodiment_manufacturing} shows the measured line roughness $R_z$ of the six spindle shafts. 
Measurements were acquired with a Keyence VHX-6000 digital microscope at 800× magnification, measured at four radial positions offset by 90° at the axial midpoint of the shaft. At each position, a $5\,\si{\milli\metre}$ evaluation profile was extracted along the shaft axis with a cut-off wavelength of $\lambda_c = 0.8\,\si{\milli\metre}$ in accordance with ISO~4288.
The mean roughness is $R_z = 2.205\,\si{\micro\metre}$ with a standard deviation of $sd = 0.171\,\si{\micro\metre}$. 
Although all values are below the specified maximum of $R_z = 2.5\,\si{\micro\metre}$, the distribution lies close to the tolerance limit.
For one component, a local roughness value of $R_z = 5.04\,\si{\micro\metre}$ is observed at a specific circumferential position. 
Post-analysis indicates a localized surface defect affecting the measurement. 
This highlights that spatial orientation is not explicitly encoded in the nominal characteristic space, which becomes relevant for later degradation description.

\subsubsection{Step 3: Detection and quantification of degradation patterns}
\begin{figure}[pos=b]
    \centering
    \includegraphics[width=16cm]{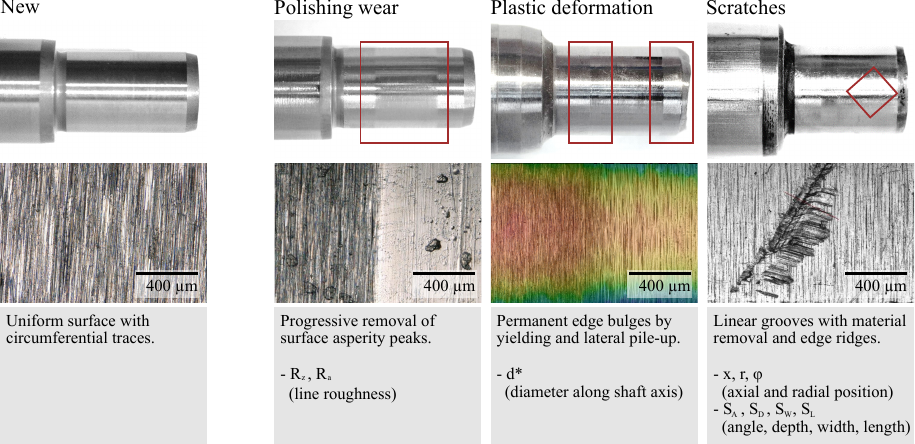}
    \caption{Observed degradation patterns of the spindle shafts. Macroscopic view (top) where red boxes mark the degradation, microscopic detail (middle) with qualitative description and measurable parameters (bottom). The new shaft serves as reference baseline. Patterns can occur individually or in combination on a single specimen.}
    \label{fig:casestudy_degradationpatterns}
\end{figure}
\begin{figure}[pos=h]
    \centering
    \includegraphics[width=15cm]{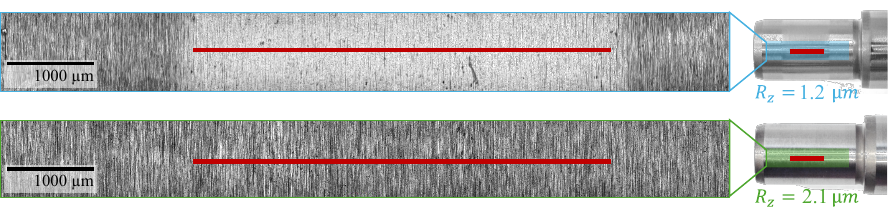}
    \caption{Surface roughness characterization of raceway acquired with an Alicona µCMM at 20× magnification. The red line indicates the 5 mm evaluation profile. Top: Used spindle with reduced surface roughness ($R_z = 1.2 \si{\micro\metre}$). Bottom: New spindle in as-manufactured condition ($R_z = 2.1 \si{\micro\metre}$).}
    \label{fig:casestudy_ploshing_quantification}
\end{figure}
\begin{figure}[pos=h]
    \centering
    \includegraphics[width=15.4cm]{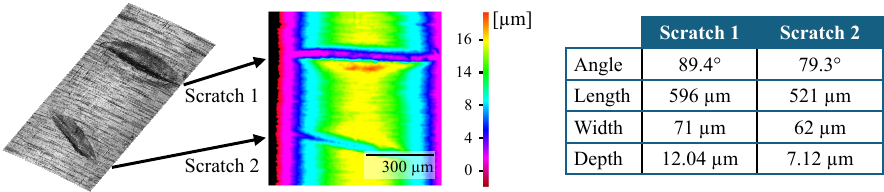}
    \caption{Exemplary quantification of a surface scratch using optical topography data. A perpendicular profile determines scratch width and depth via three characteristic points, while the visible extent in the image is measured along the scratch direction to determine its length.}
    \label{fig:casestudy_scratch}
\end{figure}

The degradation knowledge base is established by combining literature-based failure classifications with empirical observations from used spindle shafts. 
In particular, bearing-related degradation modes according to ISO 15243~\citep{bearing_damage_norm} are used as a structural reference, as the spindle shaft operates as the inner ring of the NRB.
Combined with the screening of the 21 used spindle shafts exhibiting representative degradation patterns are selected for detailed analysis. 
The screening also revealed that several shafts exhibited multiple degradation patterns simultaneously, in some cases overlapping in the same region of the raceway. 
While such combinations are inherent to real used components, they complicate the unambiguous attribution of measured deviations to a single degradation mechanism. 
The resulting knowledge base links observed degradation patterns to their embodiment manifestation and corresponding geometric descriptors, as summarized in \cref{fig:casestudy_degradationpatterns}.
\\
\textbf{Polishing wear}: Characterized by a reduction in surface roughness due to repeated contact between rollers and raceway. 
It is quantified using local roughness measurements on the shaft surface.
\textbf{Plastic deformation}: Caused by high contact stresses between roller edges and shaft surface, leading to a reduction in local diameter.
It is represented through axial diameter measurements and comparison to the nominal geometry. 
\textbf{Scratches}: Local abrasive damage caused by hard particles. 
Represented through spatial location and geometric descriptors such as angle, depth, width, and length.
\\
In this definition polishing wear as well as plastic deformation are continuous degradation patterns, while the discrete pattern of scratches needs localization of the defects.
The empirical analysis also revealed cases of fretting corrosion and pitting, which are omitted here due to scarce appearance.
\\
For the initial degradation identification, a visual pre-screening was used to identify and localize relevant degradation features in the dataset. 
Since the focus of this case study is on the embodiment-based representation rather than automated inspection strategies, the detection process is not further formalized or automated in this work.
Building on this pre-screening, the identified degradation patterns are subsequently quantified using dedicated measurement systems. 
\\
%Polishing wear is initially identified through visual inspection based on a characteristic change in surface reflectivity in industrial camera images (cf. Fig. \cref{fig:casestudy_ploshing_quantification}). 
\textbf{Polishing wear} and its associated surface roughness changes, are quantified using an Alicona \textmu CMM at $20\times$ magnification. 
A $5\,\si{\milli\metre}$ evaluation profile is extracted along the shaft axis within the identified contact zone, using a cut-off wavelength of $\lambda_c = 0.8\,\si{\milli\metre}$ in accordance with ISO~4288. 
For reference, the same measurement procedure is applied to an unused spindle shaft to enable comparison with the manufacturing-induced baseline established in step~2.
The results are shown in \cref{fig:casestudy_ploshing_quantification}. 
The worn spindle exhibits a mean roughness depth of $R_z = 1.2\,\si{\micro\metre}$ within the bearing contact zone, compared to $R_z = 2.1\,\si{\micro\metre}$ for the new spindle, corresponding to a reduction of approximately $43\,\si{\percent}$. 
This confirms that the visually identified region corresponds to a measurable modification of surface topography, consistent with the smoothing effect typically associated with polishing wear.
\begin{figure}[pos=b]
    \centering
    \includegraphics[width=11.4cm]{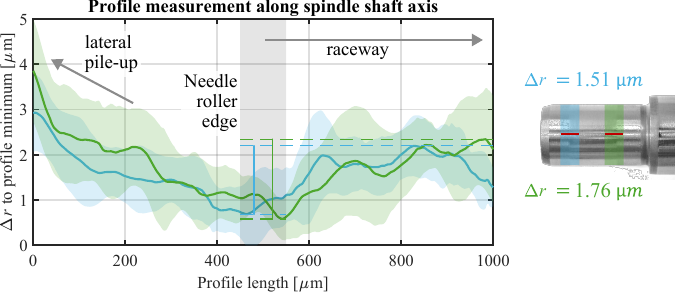}
    \caption{Surface profile along the spindle shaft axis at the contact edge. The transition towards the bevel gear (green) shows a more pronounced radial reduction of $\Delta r = 1.76 \,\si{\micro\metre}$, than the shaft-end side (blue) with $\Delta r = 1.51 \,\si{\micro\metre}$. Both profiles rise beyond the contact edge, indicating lateral pile-up of displaced material.}
    \label{fig:casestudy_plasticdefomration}
\end{figure}
\\
\textbf{Plastic deformation} is evaluated through 3D topography measurements using a Keyence VHX-6000 digital microscope at 600× magnification, from which surface profiles are extracted.
The analysis focuses on the transition regions between the bearing raceway and the nominal shaft surface, where contact-induced material flow is expected to occur.
For each transition region, four profiles distributed along the circumference are recorded, with two measurement lines per angular position ($n = 8$ per side).
The resulting radius deviations are aggregated and summarized in \cref{fig:casestudy_plasticdefomration}.
Both transition regions exhibit a local radius reduction at the needle roller edge.
$\Delta r = 1.51\,\si{\micro\metre}$ on the shaft-end side and $\Delta r = 1.76\,\si{\micro\metre}$ on the bevel-gear side.
This corresponds to a mean local diameter reduction of approximately $\Delta d \approx 3.3\,\si{\micro\metre}$.
Beyond the contact zone, both profiles rise again towards the shaft ends, consistent with a lateral material pile-up displaced from the raceway during operation.
Compared to the manufacturing-induced variability of the diameter established in step 2 (standard deviation $sd = 0.789\,\si{\micro\metre}$), the observed deviations remain within a comparable order of magnitude.
Accordingly, plastic deformation is included in the set of investigated defect classes, but does not indicate a pronounced degradation effect for the analyzed sample set.
\\
\textbf{Scratches} are quantified by local geometric descriptors derived from high-resolution topography measurements performed with the Alicona \textmu CMM. 
For each scratch, the point cloud acquired from the topographic scan is used to manually select characteristic points along and across the scratch. 
The length and width are calculated as Euclidean distances between opposing boundary points. 
The depth is derived from a roughness profile extracted perpendicular to the scratch orientation, using a cut-off wavelength of $\lambda_c = 0.8\,\si{\milli\metre}$. It is evaluated as the height difference between the local surface level and the minimum profile point within the groove. 
The orientation is defined as the angle between the scratch’s principal axis and the shaft axis, using the surface machining marks as reference.
Two scratches are analyzed on the same shaft, with their geometric descriptors summarized in \cref{fig:casestudy_scratch}. 
Scratch 1 exhibits a length of $D_L=596\,\si{\micro\metre}$, a width of $S_W=71\,\si{\micro\metre}$, a depth of $S_D=12.04\,\si{\micro\metre}$, and is nearly aligned with the circumferential direction ($S_A=89.4\,\si{\degree}$). 
Scratch 2 shows a length of $S_L=521\,\si{\micro\metre}$, a width of $S_W=62\,\si{\micro\metre}$, a depth of $S_D=7.72\,\si{\micro\metre}$, and an orientation of approximately $S_A=97.3\,\si{\degree}$ relative to the shaft axis.

\subsubsection{Step 4: Integration into the unified characteristic space}
\begin{figure}[pos=t]
    \centering
    \includegraphics[width=16cm]{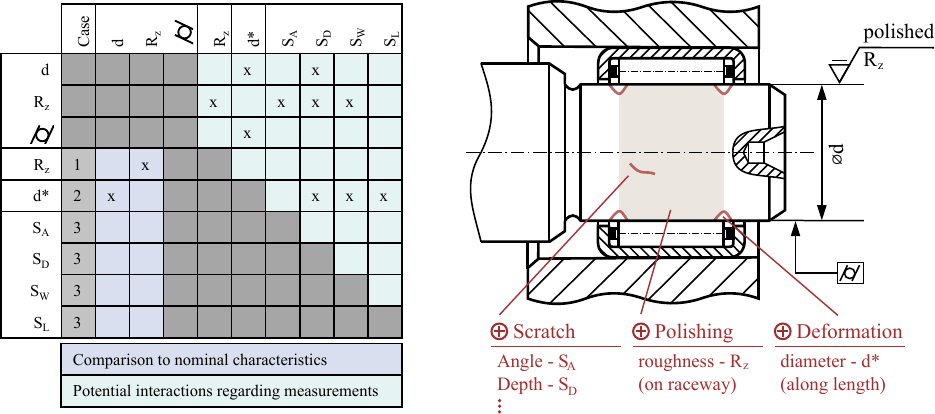}
    \caption{Integration of embodiment characteristics including comparison and potential interactions with measurements (left). Unified characteristic space (right), including the nominal characteristic space and the degradation patterns polishing wear, plastic deformation and scratches.}
    \label{fig:casestudy_unified_description}
\end{figure}

The nominal characteristics diameter, cylindricity, and surface roughness are mapped to the corresponding degradation manifestations and summarized in \cref{fig:casestudy_unified_description}.
Polishing wear is assigned to case 1, as it can be directly represented by the existing surface roughness characteristic without modification.
Plastic deformation is assigned to case 2, since the diameter is already part of the nominal characteristic space but must be extended from a discrete measurement to a continuous profile along the shaft axis to capture local reductions.
Scratches are assigned to case 3, as their occurrence introduces new, spatially localized characteristics that are not part of the original design specification.
The description is reduced to orientation, length, width and depth, as a full spatially resolved representation of individual scratch positions would lead to a prohibitively high complexity for the manual case study.
\\
In a second step, interactions between characteristics and measurement procedures are identified.
A key interaction is observed between surface roughness in the nominal characteristic space and polishing wear, requiring roughness measurements to be consistently performed within the functional raceway region, where degradation is relevant.
Plastic deformation introduces a continuous variation of diameter along the shaft axis, requiring multiple measurements along the raceway, particularly at the contact edges. 
This aligns with the measurement principle used for roundness evaluation.
Scratches affect multiple measurement outcomes depending on their orientation. 
While diameter measurements are taken circumferentially, surface roughness is evaluated along the axial direction. 
Consequently, the influence of scratches on measured characteristics depends on their spatial alignment relative to the measurement direction.
The resulting unified characteristic space, including all relevant characteristics and degradation-induced extensions, is shown on the right side of \cref{fig:casestudy_unified_description}. Deviations in geometry by scratches, polishing, and deformation are highlighted in the technical drawing with red annotations, and the corresponding embodiment characteristics are listed below the associated degradation patterns. Since each inspection yields instance-specific degradation patterns and severity, the unified characteristic space adapts dynamically as further information about the embodiment is acquired. To fully exploit the UED, a spatial representation of the embodiment, for example in a 3D point cloud, can support the localization of discrete scratches and thereby improve the spatial traceability of inspection results.
\subsubsection{Step 5: Systematic variation within the unified characteristic space}
The parameter variation, summarized in \Cref{tab:doe}, follows a staged experimental design with three phases. 
The design was adapted during execution based on intermediate results from phase 1 and 2, which motivated a focused continuation in the respective subsequent phases.
In phase 1 the characteristics diameter and surface roughness are varied using a full-factorial design with three levels each and one specimen per combination, resulting in 9 shafts with two measurement replications each.
The reference state is represented by the shaft at diameter level 0 and roughness level 0, as its configuration corresponds to diameter and surface roughness of a new component.
An increased \textbf{diameter} enlarges the contact area between shaft and needles while reducing fit clearance, thereby affecting Hertzian contact conditions.
This is expected to increase subsystem stiffness and influence vibration behavior.
The factor levels are derived from nominal design specifications and manufacturing constraints. 
Level 0 corresponds to the nominal h6 tolerance but is tightened to IT5 ($d=7h5\,\si{\milli\metre}$) for clearer differentiation between factor levels. 
Level +1 ($d=7s5\,\si{\milli\metre}$) represents an increased diameter as observed in transition fits (manufacturer D), while level -1 ($d=7e5\,\si{\milli\metre}$) reflects a reduced diameter achievable through subtractive reprocessing.
\\
\textbf{Surface roughness} influences frictional conditions at the contact interface and thereby affects the rolling behavior of the needles.
Increased roughness may lead to more localized contact and higher vibration levels, while reduced roughness may promote slip.
No significant effect on stiffness is assumed.
The factor range combines nominal design specifications, manufacturing-induced variation, and degradation effects such as polishing wear and general surface deterioration. 
Level 0 represents the nominal state ($R_z=2.5\pm0.15\,\si{\micro\metre}$), level -1 corresponds to polishing wear ($R_z=1.5\pm0.15\,\si{\micro\metre}$), and level +1 represents increased roughness due to wear-related surface damage ($R_z=3.5\pm0.15\,\si{\micro\metre}$).
Intermediate analysis of phase 1 revealed a pronounced effect of diameter on the functional responses, while no significant effect of surface roughness was observed within the investigated range. Consequently, subsequent phases were restricted to the nominal roughness level ($R_z=2.5\pm0.15\,\si{\micro\metre}$) to reduce experimental effort while maintaining a representative reference state.
\\
In phases 2 and 3, a \textbf{scratch} is introduced as a discrete degradation to capture localized disturbances expected to influence vibration behavior. 
The scratch location is fixed at the center of the raceway, and length and width are kept constant, while angle and depth are varied to represent characteristic extremes observed in degraded components.
Phase 2 addresses scratch angle as a hard-to-change factor using a split-plot design. 
The three shafts from phase 1 at the nominal roughness level are reused and complemented by three additional, nominally identical shafts, yielding two shafts per diameter level with for measurement replications for each shaft. 
Within each diameter level, one shaft is assigned to each angle ($S_A=0^\circ$ and $S_A=90^\circ$), so that angle is varied between subjects while all other characteristics remain fixed with a scratch depth of $S_D=10\,\si{\micro\metre}$.\\
Intermediate analysis of phase~2 did not reveal a resolvable distinction between axial and radial scratches at a depth of $S_D=10\,\si{\micro\metre}$, as the defect-induced excitation remained within the system-level variability. 
Consequently, phase 3 varies scratch depth within the configurations of phase 2 to capture the effect of defect severity under identical geometric and orientational conditions. 
Each axial scratch is measured at an additional depth level of $S_D=20\,\si{\micro\metre}$), providing a repeated-measures structure.
\begin{table}[pos=t]
  \centering
  \caption{Study design for functional evaluation. Phase~1 (full factorial) varies the 
           shaft diameter $D$ and the surface roughness $R_z$. 
           Phase~2 (between shaft) and phase~3 (within shaft) vary the scratch angle $S_A$ 
           and scratch depth $S_D$ within a single shaft.}
  \label{tab:doe}
  \sisetup{
    separate-uncertainty = true,
    table-align-uncertainty = true,
  }
  \setlength{\tabcolsep}{8pt}
  \begin{tabular}{c c c c S[table-format=1.2(2)] c c}
    \toprule
    & \multicolumn{4}{c}{\textbf{Phase 1: full factorial}} 
    & \multicolumn{1}{c}{\textbf{Phase 2: between shaft}} 
    & \multicolumn{1}{c}{\textbf{Phase 3: within shaft}} \\
    \cmidrule(lr){2-5} \cmidrule(lr){6-6} \cmidrule(lr){7-7}
    \textbf{Shaft ID} 
    & {Level} & {$d$ in \si{\milli\metre}} 
    & {Level} & \multicolumn{1}{c}{$R_\mathrm{z}$ in \si{\micro\metre}} 
    & {$S_\mathrm{A}$ in \si{\degree}} 
    & {$S_\mathrm{D}$ in \si{\micro\metre}} \\
    \midrule
    \textbf{1}   & $-1$           & 7e5 & $-1$           & 1.50 \pm 0.15 & --  & --      \\
    \textbf{2.1} & $-1$           & 7e5 & \phantom{$-$}0 & 2.50 \pm 0.15 & 0   & 0 / 10 / 20 \\
    \textbf{2.2} & $-1$           & 7e5 & \phantom{$-$}0 & 2.50 \pm 0.15 & 90  & 0 / 10  \\
    \textbf{3}   & $-1$           & 7e5 & $+1$           & 3.50 \pm 0.15 & --  & --      \\
    \midrule
    \textbf{4}   & \phantom{$-$}0 & 7h5 & $-1$           & 1.50 \pm 0.15 & --  & --      \\
    \textbf{5.1} & \phantom{$-$}0 & 7h5 & \phantom{$-$}0 & 2.50 \pm 0.15 & 0   & 0 / 10 / 20 \\
    \textbf{5.2} & \phantom{$-$}0 & 7h5 & \phantom{$-$}0 & 2.50 \pm 0.15 & 90  & 0 / 10  \\
    \textbf{6}   & \phantom{$-$}0 & 7h5 & $+1$           & 3.50 \pm 0.15 & --  & --      \\
    \midrule
    \textbf{7}   & $+1$           & 7s5 & $-1$           & 1.50 \pm 0.15 & --  & --      \\
    \textbf{8.1} & $+1$           & 7s5 & \phantom{$-$}0 & 2.50 \pm 0.15 & 0   & 0 / 10 / 20 \\
    \textbf{8.2} & $+1$           & 7s5 & \phantom{$-$}0 & 2.50 \pm 0.15 & 90  & 0 / 10  \\
    \textbf{9}   & $+1$           & 7s5 & $+1$           & 3.50 \pm 0.15 & --  & --      \\
    \bottomrule
  \end{tabular}
\end{table}

\subsubsection{Step 6: Quantification and evaluation of embodiment-function relations}
\begin{figure}[pos=t]
    \centering
    \includegraphics[width=0.9\linewidth]{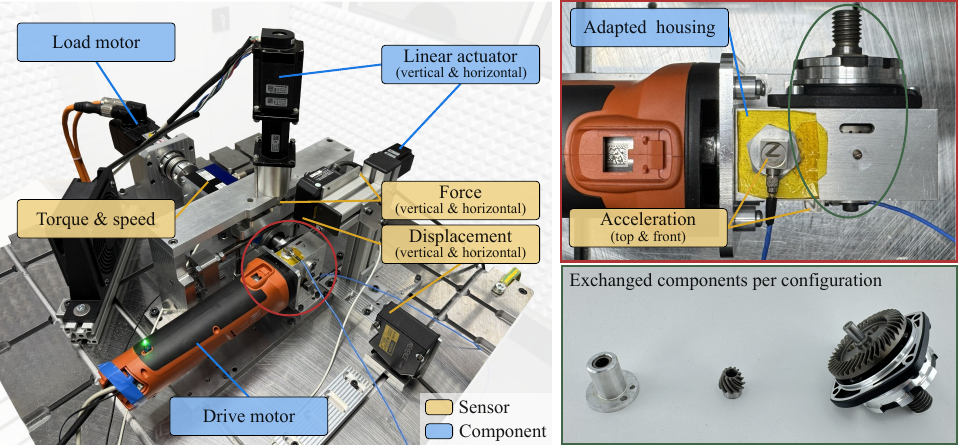}
    \caption{Test bench for investigating the influence of the spindle shaft on stiffness and vibration. Instrumented drive train loaded with torque and forces (left), adapted gearbox housing integrating original angle grinder components (top), and exchanged shaft-bearing subsystem (bottom).}
    \label{fig:casestudy_behavior_tesbench}
\end{figure}

\begin{figure}
\centering
\includegraphics[width=15.6cm]{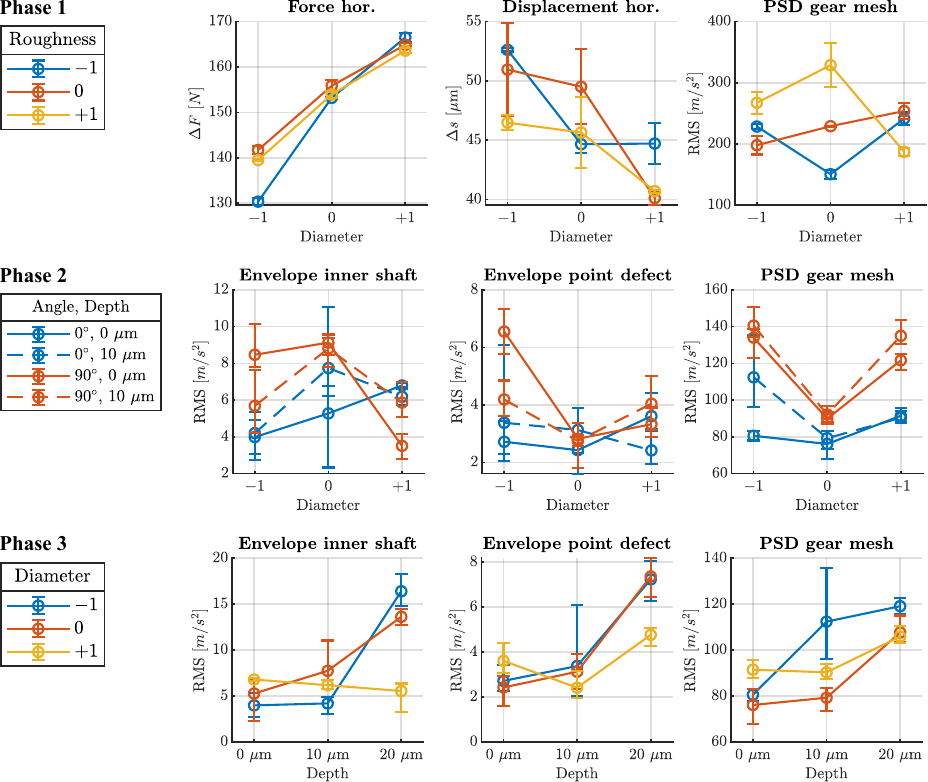}
\caption{Functional evaluation of shaft characteristics on subsystem stiffness and vibration across three phases. Phase~1 (top) evaluates horizontal force, axial bevel-gear displacement, and GMF-band RMS against shaft diameter and roughness. Phase~2 (center) and Phase~3 (bottom) assess envelope-based defect metrics and GMF-band RMS for shallow scratches angle and depth in combination with shaft diameter. Error bars indicate the range across repeated measurements.}
\label{fig:influence_degradation_overall}
\end{figure}

To investigate the effects of the characteristic variations, empirical testing is required, as no sufficiently detailed simulation model is available and domain-specific knowledge on angle grinder spindle shafts is limited in the literature.
Test specimens are therefore generated via controlled parameter variation, and all measurements are conducted on a dedicated experimental test bench.
The aim of the following experiments is to identify quantitative relations between embodiment characteristics and functional behavior, not to provide a comprehensive statistical analysis.
The number of repetitions per configuration is deliberately limited, and the focus lies on the comparative trends across factor levels rather than on absolute values or significance testing.
Reported error bars indicate the range across repeated measurements and serve to illustrate the order of magnitude of measurement scatter relative to the observed effects.
\\
The test bench is shown in \cref{fig:casestudy_behavior_tesbench} and described in more detail in \cite{zotero-item-1805}.
The original drivetrain of the angle grinder serves as power input, while a load motor and linear actuators emulate operational torque and forces.
Most original components are retained to preserve realistic system behavior. 
The gearbox housing, however, is modified to enable fast exchange of the NRB and the spindle shaft subsystem, comprising the spindle shaft, bearing and bevel gear stage.
This subsystem-level exchange is necessary because press fits and wear-related constraints in the gear stage prevent isolated shaft replacement.
To ensure reproducibility across repeated assembly cycles, bearing lubrication is omitted.
\\
The load profile is a steady operating state representing a cutting operation. 
A load motor torque of $1.5\,\mathrm{Nm}$, a horizontal actuator displacement of $200\,\si{\micro\metre}$, and a vertical actuator displacement held at $0\,\si{\micro\metre}$ are specified.
The drive motor maintains an average speed of $6400\,\si{\per\minute}$ through active control.
Functional behavior is quantified using dedicated metrics and measurement equipment, with all signals sampled at $18\,\si{\kilo\hertz}$.
Torque and rotational speed are captured by a combined measurement shaft (HBK T210).
Stiffness is derived from the reaction force measured at the actuator by a strain-gauge load cell (Burster 8435-6001), complemented by bevel gear displacement, measured in two orthogonal directions with laser sensors (Keyence LK-H052 \& LK-H152) as an indirect indicator of shaft deflection.
Vibration emissions are captured by two accelerometers (PCB 356B21) mounted on the front and top faces of the gearbox housing.
Phase~1 uses the vertical component of the top-mounted accelerometer, while phases~2 and~3 use the axial component of the front-mounted accelerometer.
Acceleration signals are analyzed in the frequency domain using the power spectral density (PSD), which describes the distribution of signal energy across frequency.
From the PSD, the root-mean-square (RMS) value within a defined frequency band is computed as a scalar measure of vibration intensity in that band.
For broadband vibration, the RMS in the gear-mesh frequency (GMF) band is used.
For bearing-related effects, the envelope of the acceleration signal is computed and the RMS is evaluated around two characteristic kinematic frequencies of the NRB.
The inner-race rotational frequency indicates eccentric loading of the shaft and the inner-race point-defect frequency indicates localized damage on the shaft surface.
Results for the three phases are summarized in \cref{fig:influence_degradation_overall}, where error bars indicate the range across repeated measurements.
\\
\\
\textbf{Phase 1 -- Diameter and roughness.}\\
The first phase investigates the influence of shaft diameter and surface roughness in the absence of scratches.
The horizontal force exhibits a clear dependency on shaft diameter, with three distinct plateaus corresponding to the three diameter levels and indicating a systematic increase in subsystem stiffness with diameter.
The axial displacement of the bevel gear confirms this trend as displacement decreases at higher stiffness, again well separated across diameter levels.
In contrast, variations in surface roughness produce only minor differences that remain within the scatter of repeated measurements.
The broadband vibration metric in the GMF band shows no systematic trend with respect to either diameter or roughness.
Phase~1 therefore establishes diameter as the dominant driver of subsystem stiffness, while neither diameter nor roughness produces a resolvable effect on vibration within the investigated range.
This negative result on the vibration side also confirms that the broadband vibration baseline is sufficiently clean to be used as a detector for localized defects in the subsequent phases.
\\
\textbf{Phase 2 -- Scratch angle.}\\
The second phase introduces a shallow scratch of approximately $S_D=10\,\si{\micro\metre}$ depth and varies its orientation between axial ($S_A=0^{\circ}$) and radial ($S_A=90^{\circ}$).
The intent is to probe whether scratch orientation can be discriminated through the vibration response and whether the bearing-specific envelope metrics provide a more sensitive indicator than the broadband GMF-band RMS.
Across all three vibration metrics, no systematic separation between scratch orientations can be resolved.
A notable observation is that the reference measurements (solid line) taken before scratching the shaft already differ between configurations that should be nominally identical.
In contrast, the post-scratch measurements (dashed line) remain close to their respective references.
This pattern indicates that the variability is not driven by random measurement noise of the test bench, but rather by systematic effects associated with assembly and the remaining components of the subsystem.
At this scratch depth, the defect-induced excitation is therefore of the same order as the global system variability, and a reliable distinction between axial and radial scratches is not yet possible.
\\
\textbf{Phase 3 -- Scratch depth.}\\
Building on the findings of phase~2, the third phase isolates the effect of scratch depth by restricting the analysis to axial scratches and extending the depth range up to $S_D=20\,\si{\micro\metre}$.
With increasing scratch depth, all three vibration metrics show a consistent upward trend.
The effect is most pronounced for the envelope-based metrics, where the response at the deepest scratch level clearly exceeds both the reference and the shallow-scratch level.
The broadband GMF-band RMS show a comparable, although less sharply separated, increase with scratch depth.
The progression across the three depth levels is markedly non-linear.
A difference between the reference state and the $S_D=10\,\si{\micro\metre}$ level is small and remains within the scatter observed in phase~2, whereas the step to $S_D=20\,\si{\micro\metre}$ produces a substantial rise across all metrics.
This indicates the existence of a detection threshold below which axial scratches cannot be reliably resolved by the considered vibration features, and above which the defect emerges clearly from the system-level variability.
An interaction with the variation in shaft diameter could not be detected.

\subsubsection{Step 7: Definition of functionally derived tolerance regions}
Building on the quantified relations, tolerance regions for used shafts can be derived under a set of assumptions. 
Diameter is treated as the dominant driver of stiffness and axial scratches as the dominant driver of vibration once a sufficient depth is reached.
Radial scratch orientation is neglected as no clear effect on vibration metrics could be shown within the given scratch depths.
Since roughness does not exhibit a detectable effect on functional behavior within the examined parameter space, it is not considered as a criterion for evaluation.
However, a potential influence on long-term reliability cannot be excluded based on the present data and therefore remains subject to future investigation.
\\
For the diameter, the functional requirement is defined such that a reconditioned shaft must not exhibit lower stiffness than the original component. 
This requirement translates into a minimum admissible diameter within the h5 tolerance band of the new part. 
Shafts falling below this threshold are not discarded but can be reprocessed via additive reprocessing, thereby remaining within the circular use cycle.
The effect of diameters exceeding the h5 band on long-term reliability has not yet been quantified and requires further study.
\\
With respect to surface damage, the experimental results suggest a threshold-type behavior for axial scratches.
Shallow scratches up to approximately $S_D=10\,\si{\micro\metre}$ remain within the system-level variability and cannot be reliably distinguished from the undamaged reference, while deeper scratches in the order of $S_D=20\,\si{\micro\metre}$ produce a clear and consistent increase across all vibration metrics.
A maximum admissible scratch depth exceeding the identified threshold can therefore be defined within this range, beyond which the allowable vibration level is exceeded.
The exact derivation of the admissible scratch depth through combination of the individual vibration metrics will be part of future research.
\\
Overall, this results in a concise inspection logic for the considered circular economy scenario.
For the inspected parameter space, roughness is not used as a criterion.
Scratches are evaluated against a defined depth limit, with the axial orientation treated as the dominant mode.
The shaft diameter is assessed against the h5 tolerance, with sub-threshold components routed to additive reprocessing.

\section{Discussion}
\label{sec:discussion}
Through the proposed approach and the subsequent case study, the research question \textit{How can the superimposed embodiment of a component be described in a way that enables systematic quantification of its functional behavior for decision-making in circular manufacturing systems?} can be answered as follows:\\
The UED establishes as a basis for decision-making in circular manufacturing systems by systematically combining embodiment description approaches from design, manufacturing, and degradation.
It is structured into two coupled layers that capture and relate lifecycle-induced effects, constructed through a supporting UED method.

\subsection{The UED and its supporting UED method}
As a representation, the UED enables a structured description of the individually degraded embodiment of a component.
It meets the requirements of circular manufacturing systems through adaptation and extension of the characteristic space, which in linear production is described in a fixed manner by models such as GPS \citep{DINISO1101} or GeoSpelling \citep{dantanGeometricalProductSpecifications2008}.
This adaptation and extension is driven by the occurring manifestation of degradation instead of treating degradation as deviations within an existing parameter space or introducing additional descriptors in an ad-hoc manner, without systematically addressing their relation to the overall embodiment \citep{vanmaeleVisionassistedConditionMonitoring2025}.
The case study confirms that degradation can alter the structure of the embodiment description itself. Continuous phenomena such as polishing wear are represented within existing characteristics, plastic deformation requires adaptation of existing ones, and discrete patterns such as scratches necessitate the introduction of entirely new characteristics.
A limitation, however, stems from the definition of appropriate descriptors for degradation features.
For example, it remains unclear whether a scratch should be described by a set of parameters such as angle, length, width, and depth, or by aggregated measures such as area or volume.
The selection of suitable descriptors requires expert knowledge and may influence the resulting embodiment--function relations.
In cases where multiple degradation patterns overlap within the same region, a clear and unambiguous embodiment description could not always be established. 
This challenge becomes particularly relevant for automated quantification, and is consistent with findings in the literature that subtractive reprocessing is often required to restore a well-defined embodiment state before further evaluation \citep{barragandelosriosStudySurfaceRoughness2023}.
\\
The construction of the UED is operationalized by the supporting UED method, which derives tolerance regions for decision-making in the circular factory, such as reuse, reprocessing, or recycling of components.
While previous embodiment approaches remain at a conceptual level \citep{geroSituatedFunctionBehaviour2004} or primarily focus on ideal states \citep{jingliuSimulationAnalysisBall2023} or sensor-based monitoring \citep{liIntelligentFaultIdentification2021}, the UED method enables the identification and quantification of functionally relevant characteristics and their systematic relation to functional behavior.
The case study further indicates that the effect of individual characteristics on functional behavior cannot always be interpreted independently, as combinations such as scratch angle and depth may influence the system behavior in a coupled manner.
This underlines the necessity of considering interaction effects when establishing embodiment--function relations, and highlights that small geometric variations can have a significant impact on functional behavior, especially in high-performance systems such as bearing interfaces.
With the UED as the underlying representation, the evaluation of used components in circular factories is no longer based solely on qualitative inspection or isolated measurements, but on a model-based interpretation of their functional performance through the quantitative relations of embodiment and functional behavior.

\subsection{Positioning and limitations}
In parameter-based models for embodiment--function relations, the consideration of various individual dimensions of factor spaces causes issues with complexity, as design engineers need to decide how to model degradation patterns and features as corresponding dimensions.
These mostly implicit decisions are now supported by the structuring of characteristics into three categories and their interrelations.
In this sense, the contribution goes beyond an improved embodiment description and establishes a basis for state-dependent embodiment modeling, which is essential for decision-making in circular manufacturing contexts.
While the UED itself is not restricted to geometric characteristics and can incorporate material properties such as hardness or residual stresses through the same structuring principle, the case study deliberately focuses on geometric embodiment characteristics to maintain a clear and measurable application scope.
The generalizability of the UED to non-mechanical domains, which typically require different model types for investigation, remains to be validated.
\\
Overall, the UED can be interpreted as a bridging approach that connects design-oriented, manufacturing-oriented, and degradation-oriented embodiment descriptions.
By integrating these perspectives into an adaptive and extendable description, it provides a foundation for analyzing and interpreting real, degraded components in a systematic and function-oriented manner.
While the approach is designed to be generalizable, its applicability to other components and systems with different functional mechanisms requires further validation, as the case study is limited to the specific component of an angle grinder spindle shaft and its functional principle.
In addition, since the UED method has so far only been applied by its developers, no insights into how it is understood by other design engineers or researchers could yet be derived.
The effort required for individual method steps also varies considerably and depends strongly on the availability of technical documentation, manufacturing data, and suitable measurement and testing environments.
In particular, the identification of functionally relevant characteristics in the early method steps could be further supported by integrating established approaches such as DoE screening \citep{montgomeryDesignAnalysisExperiments2019} or the contact and channel approach \citep{graubergerContactChannelApproach2020}.
The quantification of embodiment--function relations is furthermore based on a limited parameter space and controlled experimental conditions, and real-world applications may involve more complex interactions and variability from which new challenges may emerge.
Furthermore, the application of the UED in this work is restricted to manual evaluation of functional behavior and the derivation of corresponding tolerance regions.
The linkage to automated design decision-making in manufacturing systems, which would be necessary for scalable use in operational circular factories, was not the focus of this case study and requires further investigation.

\subsection{Implications for circular manufacturing}
The UED has significant implications for decision-making and process design in circular manufacturing systems.
For diagnostics, it enables a more comprehensive and structured assessment of used components.
Instead of relying on isolated measurements or defect detection \citep{tabernikSegmentationbasedDeeplearningApproach2020, liuBevelGearQuality2016}, multiple characteristics and their interactions can be considered simultaneously, allowing a more accurate identification of degradation states and their potential impact on functional behavior.
\\
For reprocessing, the UED enables a shift from constraint-driven restoration to function-oriented modification.
Since the influence of individual characteristics on functional behavior is quantified, reprocessing strategies can be tailored to achieve specific functional targets.
This includes not only restoring nominal conditions but also intentionally modifying nominal characteristics, for example by adjusting diameters or surface properties, to improve performance or adapt components to new product generations, thereby enabling ``perpetual innovative products'' \citep{lanzaVisionCircularFactory2024}.
\\
For design, the results provide a basis for developing more robust and circularity-oriented products.
By identifying which embodiment characteristics are functionally relevant and how they interact, design engineers can derive targeted measures to reduce sensitivity to degradation or to enable more effective reprocessing.
Overall, the UED supports a transition towards model-based, function-oriented evaluation, which is essential for the implementation of circular factories.

\section{Conclusion and outlook}
\label{sec:conclusion}
Decision-making in circular manufacturing systems requires a consistent description of embodiment states that goes beyond the nominal design characteristics of linear production.
This paper introduced the UED as a two-layered, state-dependent representation of mechanical components, supported by the UED method that guides its model-building process.
The UED integrates design, manufacturing, and degradation perspectives into a unified characteristic space that adapts and extends as new lifecycle effects emerge.
It further links the resulting embodiment characteristics to the functional behavior of the surrounding subsystem through functionally derived tolerance regions.
The supporting UED method operationalizes this two-layer structure through a procedure that constructs the unified characteristic space from design and degradation information and establishes the functionally derived tolerance regions through systematic variation and testing.
\\
The case study on an angle grinder spindle shaft demonstrated the construction of the UED in a realistic circular manufacturing scenario.
Manufacturing-induced variations were established as a statistical baseline against which the degradation patterns polishing wear, plastic deformation, and scratches were systematically detected, quantified, and embedded into the unified characteristic space.
The integration revealed three distinct cases in which existing nominal characteristics either remained sufficient, required adaptation, or had to be complemented by new spatially localized descriptors.
These findings illustrate that degradation can act as a structural driver of the characteristic space rather than as a mere deviation within a fixed parameter set.
Through systematic variation and experimental evaluation on a dedicated test bench, embodiment--function relations were quantified and translated into functionally derived tolerance regions.
This enables data-driven decisions on reuse, reprocessing, and design adaptation across product generations and manufacturers.
The results also showed that the quality of the resulting UED depends on the appropriate definition of degradation descriptors, which requires domain knowledge and influences the granularity of the representation.
\\
Future research should extend the methodological support for characteristic definition and degradation modeling, for example through structured screening approaches or data-driven descriptor selection that reduce reliance on expert judgment.
The treatment of interacting or spatially overlapping degradation mechanisms remains a relevant topic for improving the robustness of the representation, particularly in cases where subtractive reprocessing is required to restore an unambiguous geometric state.
Beyond manual evaluation of functional behavior and the derivation of corresponding tolerance regions, the linkage to automated design decision-making in manufacturing systems requires further investigation, as it was not the primary focus of the present case study.
Finally, the scalability of the UED to different component classes and functional principles beyond rotational machine elements should be investigated, including the use of the supporting UED method by engineers outside the developing team to assess transferability into industrial practice.
\\
Overall, the UED provides a foundation for moving from static to state-dependent embodiment representations.
By enabling a consistent link between the physical state of components and the functional behavior of the corresponding subsystem, the UED is a central element within the control of circular manufacturing systems.
In this context, value retention and product innovation can be reconciled through informed, function-oriented decisions at the level of individual components entering a new product generation.


%\section{}\label{}

\printcredits
\\
\\
\textbf{Declaration of generative AI and AI-assisted technologies in the manuscript preparation process}\\
During the preparation of this work, the authors used ChatGPT (OpenAI) and Claude (Anthropic) in order to improve the structure, readability, and phrasing of the manuscript. 
After using these tools, the authors reviewed and edited the content as needed and take full responsibility for the content of the publication.\\
\\
\textbf{Declaration of competing interest} \\
The authors declare that they have no known competing financial interests or personal relationships that could have appeared to influence the work reported in this paper.\\
\\
\textbf{Acknowledgments} \\
Funded by the Deutsche Forschungsgemeinschaft (DFG, German Research Foundation) - SFB 1574 - 471687386.

\bibliographystyle{cas-model1-num-names-nourl}

%\bibliographystyle{cas-model2-names}

\bibliography{references}



\end{document}